\documentclass[12pt]{article}

\usepackage{amssymb}
\usepackage{amsfonts}
\usepackage{theorem}
\usepackage{graphicx}
\usepackage[leqno]{amsmath}
\usepackage{enumerate}
\usepackage{indentfirst}

\newenvironment{proof}[1][Proof]{\textbf{#1.} }{\hfill $\square$}

\newtheorem{lem}{Lemma}
\newtheorem{thm}{Theorem}
\newtheorem{defin}{Definition}

\newtheorem{prop}{Proposition}
\newtheorem{corol}{Corollary}
\newtheorem{ex}{Example}

\newcommand{\R}{\mathbb{R}}
\newcommand{\X}{\mathbb{X}}

\newcommand{\ID}{\mathbb{D}}

\newcommand{\prb}{\mathbb{P}}
\newcommand{\qprb}{\mathbb{Q}}
\newcommand{\E}{\mathbb{E}}
\newcommand{\tri}{\mathcal{F}}
\newcommand{\al}{\alpha}
\newcommand{\be}{\beta}

\newcommand{\la}{\lambda}

\newcommand{\ind}{\mathbf{1}}

\newcommand{\wtil}{\widetilde}

\newcommand{\opt}{\textrm{opt}}
\newcommand{\hdd}{\textrm{HDD}}
\newcommand{\chdd}{\textrm{cHDD}}

\title{Optimal cross hedging of insurance derivatives}
\author{Stefan Ankirchner \\ Institut f\"ur Mathematik\\
Humboldt-Universit\"at zu Berlin\\ Unter den Linden 6\\ 10099
Berlin\\ Germany \and Peter Imkeller\\ Institut f\"ur Mathematik\\
Humboldt-Universit\"at zu Berlin\\ Unter den Linden 6\\ 10099
Berlin\\ Germany \and Alexandre Popier\\Laboratoire de statistiques et processus\\
    Universit\'e du Maine\\
    Avenue Olivier Messiaen\\
    F-72085 Le Mans Cedex 9\\
    France}

\begin{document}

\maketitle

\begin{abstract}
We consider insurance derivatives depending on an external physical
risk process, for example a temperature in a low dimensional climate
model. We assume that this process is correlated with a tradable
financial asset. We derive optimal strategies for exponential
utility from terminal wealth, determine the indifference prices of
the derivatives, and interpret them in terms of diversification
pressure. Moreover we check the optimal investment strategies for
standard admissibility criteria. Finally we compare the static risk
connected with an insurance derivative to the reduced risk due to a
dynamic investment into the correlated asset. We show that dynamic
hedging reduces the risk aversion in terms of entropic risk
measures by a factor related to the correlation.
\end{abstract}



\section{Introduction}

In recent years, financial risks originating in the uncertainties of
weather and climate are gaining an increasing attention in insurance
and banking. This is partly due to the public coverage changes in
weather patterns obtain, but also to the awareness that modern
methods of stochastic finance offer new ways for dynamically
transferring insurance risk to financial markets. Not only since the
increasing activity in weather extremes such as big tropical storms
became publicly aware, there is no doubt that insurance companies,
in particular big re-insurers, have to face the dramatic effects of
climate and weather.\par\smallskip

The idea for this paper arose from discussions with a representative
of a re-insurance company that is involved in dynamical hedging of
weather risk, by offering HDD derivatives for example. These
derivatives are written on the surplus of {\em Heating Degree Days},
i.e. days on which temperatures are above the 18 degree Celsius
minimum heating temperature, and provide an instrument for hedging
the risk an energy producer faces due to a too warm heating season,
during which less energy is sold to private households. The
discussions were focussed on the idea of {\em cross hedging}, i.e.
using the potentials that are due to the negative correlation of
risk exposure of different agents on a finance and insurance market.
In a simple equilibrium model (see \cite{huimkellermueller04} and
\cite{chaumontimkellermueller05}) we previously dealt with the idea
of creating a market for hedging climate risk, for instance due to
the randomly periodic phenomenon of the sea surface temperature
anomaly of the South Pacific commonly known as El Ni\~no. To test
the concepts in a numerical simulation, we considered a set of small
financial agents composed of a fisher, a farmer and a bank, with
either negatively correlated risk exposure
or with the desire of diversification of their portfolio. The idea
of using the potentials of negative correlation of climate risk
exposure in cross hedging is of course applicable in scenarios
that are closer to the interest of re-insurers than the simple toy
example just mentioned. Just imagine energy producers in different
parts of Europe, for which the weather patterns have a strong
tendency of complementarity, due to the position of the jet stream
directing the trajectories of cyclones. And of course, re-insurers
have to be modelled as big agents, capable of influencing prices
of risky assets on a financial market designed for this
alternative risk transfer.\par\bigskip

In the simple conceptual financial market model considered in this
paper, on the one hand we try to develop a basic understanding of
the impact of the correlation of risk exposure of different agents
on prices and hedges of weather or climate derivatives. On the other
hand, we try to keep it simple enough to obtain explicit solutions
for pricing and hedging of financial products designed for climate
or weather risk, which we simply call external risk
henceforth.\par\smallskip

The basic ingredient in our model therefore is given by an
external risk process $X$ describing for example the temperature
during a heating period, the uncertainty of which is modelled by a
Wiener process $W$ with volatility function $\sigma$. We face an
agent, for instance an insurance company, who sells a derivative
which from his perspective provides the random income $F(X_T)$
depending on the position $X_T$ of the risk process at maturity
$T$. The second source of income for the agent is given by a
security on a financial market with price process $P$, which can
serve as an investment possibility for hedging the risk of the
derivative. $X$ and $P$ are correlated: the proportional
infinitesimal increment $\frac{dP_t}{P_t}$ at time $t$ contains a
component of $W$ with intensity $\beta_1^2$, a component with
intensity $\beta_2^2$ of an independent Wiener process $B$, as
well as a drift component depending on the risk process $X$. A
strategy $ \pi$ for investment into the asset leads to a wealth
process $V^{\pi}$. The agent is supposed to measure his utility
from total terminal wealth $F(X_t) + V_T^{\pi}$ with respect to an
exponential utility function $U(x) = -\exp(-\eta x),
x\in\mathbb{R},$ with risk aversion coefficient $\eta.$ In this
situation, for initial wealth $x\in\mathbb{R},$ the utility
indifference price $p$ for the derivative $F(X_T)$ is given
according to
\begin{equation}\label{utilityindifference} \sup_{\pi}
E(U(V_T^{\pi} + F(X_T) - p) = \sup_{\pi} E(U(V_T^{\pi})).
\end{equation}
Its price $p(t,x)$ at time $t\in [0,T]$ for initial wealth $x$ is
explicitly found to be
$$p(t,x) = -\frac{1}{k} \ln E^\qprb(\exp(-k F(Y_T^{t,x}))),
$$
where $\qprb$ is an equivalent probability measure and $Y^{t,x}$ is a diffusion process starting at $x$ at time $t$
and originating in the Feynman-Kac formula, and
$$k = \eta \frac{\beta_2^2}{\beta_1^2+\beta_2^2}.$$ By equally elementary and
explicit computations, formulas for the optimal investment $\pi^*$
(in the presence of the derivative) on the left hand side and
$\pi^\sharp$ (without the derivative) on right hand side of
(\ref{utilityindifference}) are obtained. For obtaining explicit
results of this type one starts with the observation that a
multiplicative decomposition of utility related to $V^{\pi}$ on
the one hand and to $F(X_T)$ on the other hand allows to handle
the associated HJB equation. This observation was already made in
Musiela and Zariphopoulou \cite{muszar},  where in the special case
of geometric $X$ and $S$ corresponding formulas were derived,
however starting from an entirely different paradigm. For a numerical analysis see \cite{monoyios}. The results
apply to provide an intriguing explicit relationship between
$\pi^*$ and $\pi^\sharp$: their difference is given by a quantity
we call {\em diversification pressure}
$$\delta(t,x) = - \beta_1 \sigma(t,x) \frac{\partial
p(t,x)}{\partial x},$$ multiplied by
$\frac{1}{\beta_1^2+\beta_2^2}$. This name is justified for
obvious reasons. Suppose that the risk sensitivity $\frac{\partial
p(t,x)}{\partial x}$ of the price is positive. If the correlation
$\beta_1 \sigma(t, x)$ between external risk and price is
negative, then the derivative $F(X_T)$ diversifies the portfolio
risk, and more will be invested in the asset, i.e.\ $\pi^* >
\pi^\sharp$. Conversely, if $\beta_1 \sigma(t, x)> 0$, then the
derivative $F(X_T)$ amplifies the financial risk, and thus $\pi^*
< \pi^\sharp$. 
\par\smallskip

We then consider two special scenarios for the market, both
interesting from our perspective of weather and climate risk: the
{\em non-degenerate case} in which the volatility function
$\sigma$ of the risk process is uniformly elliptic, and the {\em
geometric case} where volatility and drift depend linearly on the
distance $x$ from 0 and keep the diffusion on one half line. In
both cases, we are mainly interested in properties of the
diversification pressure such as finiteness, boundedness and
behavior at infinity (Theorems \ref{properties_nondeg} and
\ref{properties_geom}), and ultimately in a comparison of the
risks of {\em static and dynamic hedging}. We compare the price
$s$ of the derivative obtained without dynamic investment on the
financial asset of the market with the dynamic cross hedging
indifference price $p$. Typically, we find that cross hedging
reduces the risk aversion by the factor
$\frac{\beta_2^2}{\beta_1^2+\beta_2^2}$ (Theorem \ref{090805-1}),
viewed from the perspective of entropic risk measures which in a
natural way measure risky positions for exponential utilities. If
the correlation between external risk and security price enters
through its drift instead of its uncertainty component, we derive
equally explicit alternative calculations describing both utility
indifference price and optimal strategies. In this situation we
see that the dynamic hedging price is higher than the static price
iff the correlation of the derivative and the expected wealth from
investment into the security are negative (Theorem \ref{drift}).
Since admissibility of optimal trading strategies is not granted
in general (see Schachermayer \cite{schach01}), we give this
aspect a detailed treatment in Theorems \ref{040805-1},
\ref{quasiadmi} and \ref{quasiadmi_drift}.\par\bigskip

Here is a short outline of the presentation of the material. In
Section \ref{model} we introduce the model, formulate the optimal
cross hedging problem, describe an explicit solution of the
related HJB equation, and thus derive formulas for indifference
prices and optimal strategies. Since the solution of a control
problem obtained by Musiela and Zariphopoulou in \cite{muszar}
(see also \cite{davis2} for a similar result solved via
duality methods) covers a special case of our setting, our short
presentation is focussed on the main arguments. In the Sections
\ref{case1} and \ref{case2} we separately discuss the
non-degenerate and geometric cases, provide sufficient conditions
for the regularity of the solutions of our control problem, and
describe properties of the diversification pressure. In Section
\ref{admissibility} we check the admissibility of optimal cross
hedging strategies. After this in Section \ref{dynamic_static} we
study the difference between the static risk connected with an
insurance derivative and the reduced risk due to a dynamic
investment into the correlated asset, in terms of entropic risk
measures.

\section{The model}\label{model}

The pay-off of an insurance derivative is based on external risks
such as weather and climate. We consider here financial instruments
derived from an external risk which can be modelled as a diffusion
\begin{eqnarray} \label{weather}
dX_{t} & = & b(t,X_{t}) dt + \sigma (t,X_{t}) dW_{t} \ \mbox{and} \ X_{0} = x_{0},
\end{eqnarray}
where $W$ is a Brownian motion on a filtered probability space
$(\Omega, \mathcal F, (\mathcal F_t)_{t\ge 0}, \prb)$. Let $F$ be a
measurable and real-valued function and let $T > 0$. Then the random
variable $F(X_T)$ is an insurance derivative with maturity $T$.
Typical examples are so-called weather derivatives which are often
based on the average temperature during a certain time period at a
fixed location.

We suppose that there exists a financial market consisting of one non-risky asset with zero interest and one risky asset whose price process
is correlated to the process $X$. More precisely, we assume that the
price dynamics of the risky asset satisfies
\begin{eqnarray}\label{asset}
dP_{t} & = & P_{t} \left( g(t,X_t) dt + \be_{1} dW_{t} + \be_{2} dB_{t} \right), \ \mbox{and} \ P_0 > 0,
\end{eqnarray}
where $g :[0,T] \times \R \to \R$ is measurable, $\beta_1 \in \R$, $\beta_2 \not= 0$ and $B$ is a further Brownian motion independent of $W$. We denote by $\beta = \sqrt{\be_1^2 + \be_2^2}$ the {\em volatility} of $P$. Notice that $W$ is a Brownian motion driving both the risk process $X$ and the price process $P$, and that the risk process affects the drift part of the price process.

By an {\em investment strategy,} or simply strategy, we mean any
$(\tri_t)_{t \geq 0}$-predictable process $\pi$ with $\pi_0= 0$ and
such that the stochastic integral process of $\pi$ with respect to
$$\int_0^\cdot [ g(t,X_t) dt + \beta_1 dW_t + \beta_2 dB_t]$$
exists on $[0, T]$. We will interpret $\pi$ as the value of the portfolio fraction invested in the asset. This means that $\pi_t = P_t \theta_t$, where $\theta_t$ is the number of asset shares in the portfolio at time $t$. The wealth process resulting from an investment strategy $\pi$ is then given by
\begin{eqnarray}  \label{wealth}
dV^{\pi}_{t} & = & \pi_{t} \left( g(t,X_t) dt + \be_{1} dW_{t} + \be_{2} dB_{t} \right) \ \mbox{and} \ V^{\pi}_{0} = v_{0},
\end{eqnarray}
where $v_0$ is the initial wealth.

Suppose that an investor has preferences determined by an exponential utility function $U(v) = - \exp (- \eta v)$, with $\eta > 0$, and assume that he aims at maximising the expected value of his wealth at time $T$. If the investor has a derivative $F(X_T)$ in his portfolio, then the optimal strategy $\pi^\opt$ is determined by
\begin{equation}\label{utsup} 
\E \left[ U \left( V^{\pi^{\opt}}_{T} + F(X_{T}) \right) \right] = \sup_{\pi} \E \left[ U \left( V^{\pi}_{T} + F(X_{T}) \right) \right], 
\end{equation}
where the sup on the right hand side of (\ref{utsup}) is taken over all admissible strategies. For convenience we recall the definition.
Let $a \ge 0$. A strategy $\pi$ is said to be $a$-{\em admissible} if for all $t \in [0, T]$ we have $V^\pi_t \ge -a$, almost surely. $\pi$
will be called {\em admissible} if there exists an $a\ge 0$ such that $\pi$ is $a$-admissible.

If the utility function is finitely valued for all $x \in \R$,
then in general the optimal strategy solving the classical utility
maximization problem is not admissible (see \cite{schach01}).
However, it may be that the optimal strategy can be approximated
by admissible strategies. In this spirit the following definition
is useful.
\begin{defin}
A strategy $\pi$ is called {\ quasi-admissible} with respect to
$U$ if there exists a sequence $(\pi_n)$ of admissible strategies
such that $U(V^{\pi_n}_T)$ converges to $U(V^{\pi}_T)$ in $L^1$ as
$n\to\infty.$
\end{defin}
Observe that doubling strategies are not
quasi-admissible.\par\medskip
We aim at deriving explicitly the optimal investment strategy
$\pi^\opt$. Moreover, we want to find the {\em indifference price}
$p$ such that
\begin{equation} \label{price}
\sup_{\pi} \E \left[ U \left( V^{\pi}_{T} + F(X_{T}) - p \right) \right] = \sup_{\pi} \E \left[ U \left( V^{\pi}_{T} \right) \right].
\end{equation}
We will denote by $\pi^\sharp$ the strategy optimizing the right hand side of (\ref{price}), whereas the strategy optimizing the left hand side such that (\ref{price}) holds will be denoted by $\pi^*$.

We start by solving a 2-dimensional stochastic control problem with the two processes $X$ and $V^\pi$.

\subsection{Solving the control problem}

We compute the two sides of (\ref{price}). We make a little modification on $X$ and $V^{\pi}$ and define for all $(t,x,v) \in [0,T] \times \R^{2}$:
\begin{eqnarray}  \label{wealthattimet}
\forall r \geq t, \ V^{\pi,t,x,v}_{r} & = & v + \int_{t}^{r} \pi_{s} \left( g(s,X^{t,x}_s)  ds + \be_{1} dW_{s} + \be_{2} dB_{s} \right); \\ \label{weathertimet}
\forall r \geq t, \ X^{t,x}_{r} & = & x + \int_{t}^{r} b(s,X^{t,x}_{s}) ds + \int_{t}^{r} \sigma (s,X^{t,x}_{s}) dW_{s}.
\end{eqnarray}
We assume that there exists $\mu \in \R$ and $K \geq 0$ s.t.
\begin{itemize}
\item $\forall t \in [0,T]$, $\forall (x,x') \in \R^{2}$:
\begin{eqnarray*}
&& (x-x')( b(t,x) - b(t,x') ) \leq \mu |x-x'|^{2}, \\
&& |g(t,x)-g(t,x')| + |\sigma (t,x) - \sigma (t,x')| \leq K  |x-x'|.
\end{eqnarray*}
\item $\forall t \in [0,T]$, $\forall x \in \R$, $|g(t,x)| + |b(t,x)| + |\sigma (t,x)| \leq K (1+|x|)$.
\item $\forall t \in [0,T]$, the function $x \mapsto b(t,x)$ is continuous.
\end{itemize}
Then the SDE (\ref{wealthattimet}) and (\ref{weathertimet}) have a
unique strong solution $V^{\pi,t,x,v}$ and $X^{\pi,t,x}$.

Let us define $K^G : [0,T] \times \R \times \R \to \R$ by:
$$K^G(t,x,v) =  \sup_{\pi} \E \left[ U \left( V^{\pi,t,x,v}_{T} + G(X^{t,x}_{T}) \right) \right],$$
where $G$ is a given function. In (\ref{price}), $G$ is equal to 0 or to $F(.)-p$.

For this control problem, the Hamilton-Jacobi-Bellman equation leads
to the following (see for example \cite{FS}): for all $(t,x,v) \in
[0,T[ \times \R \times \R$
\begin{eqnarray} \nonumber
&& \frac{\partial h}{\partial t} + \frac{\sigma^{2}(t,x)}{2} \frac{\partial^{2} h}{\partial x^{2}} + b(t,x) \frac{\partial h}{\partial x} \\ \label{HJB}
&& \qquad + \sup_{\pi} \left\{ \frac{\pi^{2} \be^{2}}{2} \frac{\partial^{2} h}{\partial v^{2}} + \pi \be_{1} \sigma(t,x) \frac{\partial^{2} h}{\partial v \partial x} + \pi g(t,x) \frac{\partial h}{\partial v} \right\} = 0.
\end{eqnarray}
And $h$ satisfies
\begin{equation*}
h(T,x,v) = U (v + G(x)).
\end{equation*}
Remark that the supremum in (\ref{HJB}) is finite if $\displaystyle
\frac{\partial^{2} h}{\partial v^{2}} < 0$. We want to find an
explicit solution $h$ for (\ref{HJB}) and then by means of a
verification theorem we will show that $h = K^G$.

Since $U(v) = - \exp (- \eta v)$, the utility of the whole portfolio
is the product of the utility of the derivative and the utility
arising from the investment. Motivated by this multiplicativity
property we search $h$ such that
\begin{equation*}
h(t,x,v) = U(v) \exp \left( - \eta \Gamma (t,x) \right).
\end{equation*}
Then the condition $\displaystyle \frac{\partial^{2} h}{\partial v^{2}} < 0$ holds, $\Gamma(T,x) = G(x)$, and
\begin{eqnarray*}
&& \frac{\partial \Gamma}{\partial t} + \frac{\sigma(t,x)^{2}}{2} \frac{\partial^{2} \Gamma}{\partial x^{2}} + b(t,x) \frac{\partial \Gamma}{\partial x} - \frac{\sigma(t,x)^{2} \eta }{2} \left( \frac{\partial \Gamma}{\partial x} \right)^{2} \\
&& \qquad - \inf_{\pi} \left\{ \frac{\pi^{2} \be^{2} \eta }{2} + \pi
\be_{1} \sigma(t,x) \eta \frac{\partial \Gamma}{\partial x} - \pi
g(t,x) \right\} = 0.
\end{eqnarray*}
The optimal strategy is given by
\begin{equation} \label{optimal1}
\pi^{*} = \pi^{*}(t,x) = \frac{1}{\be^{2}} \left[
\frac{g(t,x)}{\eta} - \be_{1} \sigma(t,x) \frac{\partial
\Gamma}{\partial x} (t,x) \right],
\end{equation}
and hence we obtain
\begin{eqnarray} \nonumber
&& \frac{\partial \Gamma}{\partial t} + \frac{\sigma(t,x)^{2}}{2} \frac{\partial^{2} \Gamma}{\partial x^{2}} - \frac{\be_{2}^{2}}{\be^{2}} \frac{\sigma(t,x)^{2} \eta }{2} \left( \frac{\partial \Gamma}{\partial x} \right)^{2} \\ \label{HJB2}
&& \qquad + \left[ b(t,x) - \frac{\al \be_{1} \sigma(t,x) }{\be^{2}} \right] \frac{\partial \Gamma}{\partial x} + \frac{g(t,x)^{2}}{2 \eta \be^{2}} = 0.
\end{eqnarray}
Let $\psi$ be the function defined by
\begin{equation*}
\psi(t,x) = \exp \left( - \frac{\eta \be_{2}^{2}}{\be^{2}}
\Gamma(t,x) \right) \Longleftrightarrow \Gamma(t,x) = -\frac{\ln
\psi(t,x)}{k}, \ \mbox{with} \ k = \frac{\be_{2}^{2}}{\be^{2}} \eta.
\end{equation*}
Then
$$\psi(T,x) = \exp \left( -k G(x) \right),$$
and Equation (\ref{HJB2}) leads to
\begin{equation} \label{linpde1}
\frac{\partial \psi}{\partial t} + \frac{\sigma(t,x)^{2}}{2} \frac{\partial^{2} \psi}{\partial x^{2}} + \left[ b(t,x) -  \frac{\be_{1} g(t,x) \sigma(t,x)}{\be^{2}} \right] \frac{\partial \psi}{\partial x} - \frac{k g(t,x)^{2}}{2\be^{2} \eta} \psi = 0.
\end{equation}
Under some regularity conditions on $G$, we can apply the
Feynman-Kac formula in order to obtain a solution of the previous
PDE. We will provide sufficient conditions in later chapters, but
for the moment we simply suppose that we may do so. Then a solution
of (\ref{linpde1}) is given by
\begin{eqnarray*}
\psi^G(t,x) & = & \E \left[ \exp \left( -k G(Y^{t,x}_{T}) \right) \exp \left( - \frac{k}{2\be^{2} \eta } \int_{t}^{T} g(r,Y^{t,x}_r)^{2} dr \right) \right],
\end{eqnarray*}
where $Y^{t,x}$ is the solution of the following SDE:
\begin{equation}\label{crossSDE}
Y^{t,x}_{r} = x + \int_{t}^{r} \widehat{b} (s,Y^{t,x}_{s}) ds + \int_{t}^{r} \sigma (s, Y^{t,x}_{s}) d \widehat W_{s},
\end{equation}
where $\displaystyle \widehat{b}(t,x) = b(t,x) - \frac{\be_{1}
g(t,x) \sigma(t,x) }{\be^{2}}$, and $\widehat W$ is an arbitrary
one-dimensional Brownian motion. As a consequence the solution of
(\ref{HJB}) is given by
\begin{eqnarray} \nonumber
h(t,x,v) & = & U(v) \psi^G(t,x)^{\eta/k} \\ \label{HJBsol}
& = & U(v) \left\{ \E \left[ \exp \left( -k G(Y^{t,x}_{T}) \right) \exp \left( - \frac{k}{2\be^{2} \eta } \int_{t}^{T} g(r,Y^{t,x}_r)^{2} dr \right) \right] \right\}^{\eta/k}.
\end{eqnarray}
If for any $t$ and $x$ the stochastic integral $(\frac{\beta_1}{\beta^2}  g(\cdot,X^{t,x})\cdot W)$ satisfies Novikov's condition, then we can define a new probability measure $\widehat \prb$ with density
\begin{equation}\label{equivalentprb}
\frac{d \widehat \prb}{d \prb} = \exp \left( - \int_t^{T-t} \frac{\beta_1 g(s,X^{t,x}_s)}{\beta^2} dW_s - \frac12 \int_t^{T-t} \frac{\beta^2_1 g^2(s,X^{t,x}_s)}{\beta^4} ds \right).
\end{equation}
Note that the distribution of $Y^{t,x}$ relative to $\prb$ coincides with the distribution of $X^{t,x}$ relative to $\widehat \prb$, and hence in this case we may write $\psi^G$ in terms of $X^{x,t}$ and $\widehat \prb$, and
\[ h(t,x,v) = U(v) \left\{ \E^{\widehat \prb} \left[ \exp \left( -k G(X^{t,x}_{T}) \right) \exp \left( - \frac{k}{2\be^{2} \eta } \int_{t}^{T} g(r,X^{t,x}_r)^{2} dr \right) \right] \right\}^{\eta/k}.
\]

\subsubsection*{The right hand side of (\ref{price}).}

Here we take $G=0$. Then the value function $K^0(t,x,v)$ satisfies
\begin{equation} \label{pricerhs}
K^0(t,x,v) = \sup_{\pi} \E \left[ U \left( V^{\pi,t,x,v}_{T} \right) \right],
\end{equation}
and a solution of the Hamilton-Jacobi-Bellman equation is given by
\begin{equation} \label{HJBsol1}
h^0(t,x,v) = U(v) \E \left[ \exp \left( - \frac{k}{2\be^{2} \eta } \int_{t}^{T} g(r,Y^{t,x}_r)^{2} dr \right) \right]^{\eta/k}= U(v) \psi^0(t,x)^{\eta/k}.
\end{equation}
Assume at the moment that the optimal strategy $\pi^\sharp$ is quasi-admissible and that the following property is satisfied:

\vspace{0.3cm}
\noindent \textbf{Regularity Property 1:}
\textit{there exists an open set} $\mathcal{V} \subseteq \R$ \textit{such that} $h^0$ \textit{belongs to the class} $C^{1,2}([0,T[ \times \mathcal{V} \times \R)$, \textit{and for all} $x\in \mathcal{V}$, \textit{the process} $X^{t,x}$ \textit{stays in} $\mathcal{V}$.

\vspace{0.3cm}
This means that $h^0$ is a classical solution of the HJB equation (\ref{HJB}) on $[0,T] \times \mathcal{V} \times \R$.
Using the Verification Theorem (see \cite{FS}, Chapter IV), we obtain $h^0(t,x,v) = K^0(t,x,v)$ on $[0,T] \times \mathcal{V} \times \R$. Moreover the optimal strategy for (\ref{pricerhs}) is given by
\begin{equation} \label{optimal2}
\pi^{\sharp}(t,x) = \frac{1}{\be^{2}} \left[ \frac{g(t,x)}{\eta} + \frac{\be_{1} \sigma(t,x)}{k \psi^0(t,x)} \frac{\partial \psi^0}{\partial x} (t,x) \right].
\end{equation}

Observe that if $g$ is constant equal to $\al$, we find the classical result
$$K^0(t,x,v) = K^0(t,v) = \exp \left[ - \frac{\al^{2}}{2 \be^{2}}(T-t) \right] U(v) \ \mbox{and} \ \pi^{\sharp}(t,x) = \pi^{\sharp} = \frac{\al}{\eta \be^{2}}.$$


\subsubsection*{The left hand side of (\ref{price}).}

We come now to the left hand side of (\ref{price}).
Here the value function satisfies
$$K^F(t,x,v) =  \sup_{\pi} \E \left[ U \left( V^{\pi,t,x,v}_{T} + F(X^{t,x}_{T}) - p \right) \right],$$
and a solution of the related Equation (\ref{HJB}) is
\begin{eqnarray} \nonumber
h^F(t,x,v) & = & U(v) \psi^F(t,x)^{\eta/k} \\ \label{solHJB}
& = & U(v) \E \left[ e^{kp} \exp \left( -k F(Y^{t,x}_{T}) \right) \exp \left( - \frac{k}{2\be^{2} \eta } \int_{t}^{T} g(r,Y^{t,x}_r)^{2} dr \right) \right]^{\eta/k}.
\end{eqnarray}
Only temporarily we will assume that the optimal strategy $\pi^*$ is quasi-admissible and that $h^F$ is regular in the following sense.

\vspace{0.3cm}
\noindent \textbf{Regularity Property 2:}
\textit{there exists an open set} $\mathcal{U} \subseteq \mathcal{V}$ \textit{such that} $h^F$ \textit{belongs to the class} $C^{1,2}([0,T[ \times \mathcal{U} \times \R)$, \textit{and for all} $x\in \mathcal{U}$, \textit{the process} $X^{t,x}$ \textit{stays in} $\mathcal{U}$.

\vspace{0.3cm}
This means that $h^F$ is a classical solution of the HJB equation (\ref{HJB}) on $[0,T] \times \mathcal{U} \times \R$. Therefore, by using the Verification Theorem (see \cite{FS}, chapter IV), we can prove that if $v \in \R$ and $x \in \mathcal{U}$, then
$$h^F(t,x,v) = K^F(t,x,v).$$
In Section \ref{secadmi} we will provide sufficient conditions for the optimal strategies $\pi^\sharp$ and $\pi^*$ to be quasi-admissible. In the next two sections we will develop two important cases and provide
sufficient conditions for both Regularity Property 1 and 2 to hold. 
To this end we will have to appropriately restrict the control problem
on the set $[0,T] \times \mathcal{U} \times \R$. In the remainder of
this section, however, we simply {\em assume} Regularity Property 1
and 2 to be satisfied. We can thus stay in a more general framework
and discuss general results on the indifference price and the
optimal cross hedging strategy.

\subsection{The indifference price $p$}
In this subsection we will have a closer look at the indifference
price. First note that we obtain an indifference price $p(t,x)$ at
any time $t$ and for any risk level $x$ by setting
\begin{equation} \label{dynprice}
\sup_{\pi} \E \left[ U \left( V^{\pi,t,x,v}_{T} + F(X^{t,x}_{T}) - p(t,x) \right) \right] = \sup_{\pi} \E \left[ U \left( V^{\pi,t,x,v}_{T} \right) \right].
\end{equation}
This is equivalent to $h^0(t,x,v) = h^F(t,x,v)$, which implies
\begin{eqnarray} \nonumber
p(t, x) =  - \frac{1}{k} \ln \frac{\E \left[ \exp \left( -k F(Y^{t,x}_{T}) \right) \exp \left( - \frac{\beta_2^2}{2\be^{4}} \int_{t}^{T} g(r,Y^{t,x}_r)^{2} dr \right) \right]}{\E \left[ \exp \left( - \frac{\beta_2^2}{2\be^{4}} \int_{t}^{T} g(r,Y^{t,x}_r)^{2} dr \right) \right]},
\end{eqnarray}
where $Y$ is the solution of the SDE (\ref{crossSDE}). Let $\qprb$ be the probability measure with density relative to $\prb$ given by
\begin{equation}\label{newq}
\frac{d \qprb}{d \prb} = \frac{\exp \left( - \frac{\be_2^2}{2\be^{4}} \int_{t}^{T} g(r,Y^{t,x}_r)^{2} dr \right)}{\E \left[ \exp \left( - \frac{\be_2^2}{2\be^{4}} \int_{t}^{T} g(r,Y^{t,x}_r)^{2} dr \right) \right]}.
\end{equation}
Then we obtain
\begin{eqnarray} \label{ptexpression}
p(t, x) = - \frac{1}{k} \ln \E^{\qprb} \exp \left[ -k  F \left( Y^{t,x}_{T} \right) \right].
\end{eqnarray}
In the case where $(\frac{\be_1}{\be^2}g(\cdot,X^{t,x}) \cdot W)$
satisfies Novikov's condition we get the following representation of
the indifference price.
\begin{thm} \label{priceunterqhat}
If $(\frac{\be_1}{\be^2}g(\cdot,X^{t,x}) \cdot W)$ satisfies Novikov's condition, then
\begin{equation} \label{pricenovi}
p(t, x) = - \frac{1}{k} \ln \E^{\widehat \qprb} \exp \left[ -k  F \left( X^{t,x}_{T} \right) \right],
\end{equation}
where
\begin{equation}\label{newqhat}
\frac{d \widehat \qprb}{d \widehat \prb} = \frac{\exp \left( - \frac{\be_2^2}{2\be^{4}} \int_{t}^{T} g(r,X^{t,x}_r)^{2} dr \right)}{\E \left[ \exp \left( - \frac{\be_2^2}{2\be^{4}} \int_{t}^{T} g(r,X^{t,x}_r)^{2} dr \right) \right]}.
\end{equation}
\end{thm}

{\bf Remark:}
\begin{enumerate}
\item Note that the indifference prices $p(t,x)$ do not depend on $v$, the initial wealth. Besides the prices do not depend on {\em both} Brownian motions $W$ and $B$, but only on the one driving the risk process $X$.
\item A similar respresentation as in (\ref{pricenovi}) has been obtained starting with a quite different
motivation by Musiela and Zariphopoulou (see Thm 3, \cite{muszar}).
However, the authors consider just the special case where the price
process is a geometric Brownian motion with constant drift, which
means in our context that $g$ is a constant function. Note that in
this case we have $\qprb = \prb$ and $\widehat \qprb = \widehat
\prb$.
\item It can be shown that under some regularity conditions, the indifference price $p(t,x)$ coincides with the
solution of a BSDE with quadratic growth (see for example
\cite{elkarouirouge}, \cite{huimkellermueller04}, and many others).
The advantage of the approach here is that we have a simpler
representation of the price by using a forward SDE. Therefore the
model assumptions required here are weaker. Moreover, we know more
about the structure properties of the price $p(t,x)$; for example on
the differentiability in $x$, which will be of crucial importance in
order to characterize the optimal cross hedging strategy $\pi^*$ (see Section \ref{secoptstr}).

On the other hand, the BSDE approach would allow for derivatives which do not only depend on the risk process $X$,
but also on the asset price $P$. But this kind of financial product is not in the focus of this paper.
\end{enumerate}

\subsection{The optimal cross hedging strategy}\label{secoptstr}
In this section we will show how the optimal strategies $\pi^*$ and
$\pi^\sharp$ are related. It turns out that the difference depends
essentially only on the sensitivity of the indifference price due to
changes in $X_t$.

First note that $p(t,x) = - \frac1k
\frac{\psi^F(t,x)}{\psi^0(t,x)}$. Therefore, if Regularity
Properties 1 and 2 are satisfied, then $p(t,x)$ is differentiable
with respect to the initial capital $x$. Moreover, Equation
(\ref{optimal1}) implies that for $(t,x) \in [0,T] \times
\mathcal{U}$ the optimal strategy $\pi^{*}$ is given by
\begin{eqnarray} \nonumber
\pi^{*}(t,x) & = & \frac{g(t,x)}{\eta \be^{2}} - \frac{\be_{1}}{\be^{2}} \sigma(t,x) \frac{\partial \Gamma}{\partial x} (t,x) \\ \nonumber
& = & \frac{g(t,x)}{\eta \be^{2}} + \frac{\be_{1}\sigma(t,x)}{k \be^{2}} \frac{1}{\psi^0(t,x)} \frac{\partial \psi^0}{\partial x} (t,x) - \frac{\be_{1}\sigma(t,x)}{\be^{2}} \frac{\partial p}{\partial x}(t,x),
\end{eqnarray}
whereas $\pi^{\sharp}$ is given by (\ref{optimal2}). By using some
elementary calculations we obtain the intriguingly simple formula
\begin{eqnarray}\label{optimal3}
\pi^{*}(t,x) = \pi^{\sharp}(t,x) - \frac{\be_{1}\sigma(t,x)}{\be^{2}} \frac{\partial p}{\partial x}(t,x).
\end{eqnarray}
The optimal strategy $\pi^*$ depends on the {\em risk sensitivity}
$\displaystyle \frac{\partial p(t, x)}{\partial x}$ of the
indifference price $p(t,x)$ of the derivative at time $t$ and risk
level $x$. Thus $\pi^*$ is equivalent to an optimal investment in an
asset with price $P$ where the drift is corrected by a quantity
which in a way measures the agent's proneness to diversify.
Therefore we introduce the following notion.
\begin{defin} The function defined by
$$\delta(t,x) = -  \beta_1 \sigma(t, x) \frac{\partial p(t, x)}{\partial x},\quad 0\le t\le T, x\in {\cal U}$$
is called {\em diversification pressure} at time $t$ and initial capital $x$.
\end{defin}
The motivation for this name choice is given by the following considerations. Suppose that the risk sensitivity $\frac{\partial p(t,x)}{\partial x}$ of the price is positive. If the correlation $\beta_1 \sigma(t, x)$ between external risk and price is negative, then the derivative $F(X_T)$ diversifies the portfolio risk, and more will be invested in the asset, i.e.\ $\pi^* > \pi^\sharp$. Conversely, if $\beta_1 \sigma(t, x)> 0$, then the derivative $F(X_T)$ amplifies the financial risk, and thus $\pi^*< \pi^\sharp$.

The following main formula captures in a concise way the relationship between $\pi^*$ and $\pi^\sharp$.
\begin{thm} The optimal cross hedging strategy $\pi^*$ differs from the classical optimal investment strategy $\pi^\sharp$ only by the diversification pressure; more precisely
\begin{equation} \label{020905-1}
\pi^{*}(t,x) = \pi^\sharp(t,x) + \frac{1}{\beta^2} \delta(t,x).
\end{equation}
\end{thm}
\subsection*{Two particular cases} \label{cases}

In the following sections, we study two particular cases,
underpinned by important examples of external risk sources. In the
first case, $\sigma$ is non degenerate. Therefore the open set
$\mathcal{U}$ in Assumptions 1 and 2 is given by $\R$. In the second
case, we suppose that $X$ is a geometric Brownian motion, i.e.
$b(t,x) = \mu x$ and $\sigma(t,x) = \nu x$. The corresponding open
set $\mathcal{U}$ is $\R^{*} = \R \setminus \{0\}$. But by a
symmetry argument, we restrict ourselves to $\R^{*}_{+}$. Moreover
if $X$ is for example the temperature expressed in degrees Kelvin or
the pressure at the sea surface, it is natural to assume that $X$
belongs to $\R^*$.

For both cases, we will prove sufficient conditions for the fundamental Regularity Properties 1 and 2 to hold.
Besides, we will be interested in describing properties of the optimal strategies and the diversification pressure.
Hence we will give some estimates on the $\delta$. 
\section{The non-degenerate case}\label{case1}

Throughout this section, we assume that $\sigma \sigma^{*}$ is uniformly elliptic, i.e.
\begin{equation} \label{unifellip}
\exists \la > 0, \ \forall (t,x) \in [0,T] \times \R, \ \forall \xi \in \R, \langle \xi, \sigma \sigma^{*}(t,x) \xi \rangle \geq \la |\xi|^{2}.
\end{equation}
We start with some examples of risk processes $X$ being uniformly
elliptic and on which some insurance derivatives may be based.
\begin{ex}
The principal external risk sources we are interested in originate
in weather and climate. The El Ni\~no sea surface temperature
anomaly of the South Pacific describes an external risk process the
main features of which can be modeled by various low dimensional
diffusion processes that satisfy Condition (\ref{unifellip}). Let us
briefly recall two of these, used in simulations of the market
performance in a simple model to trade El Ni\~no risk, in
\cite{chaumontimkellermueller05}.\par\medskip

The first one comes from a nonlinear two-dimensional stochastic differential equation coupling the thermocline depth in some area of the South Pacific with the sea surface temperature (see \cite{fangbarcilonwang99}). The system turns out to be an autonomous nonlinear stochastic oscillator which in some parameter regimes acts as a stochastically perturbed bistable differential equation with an intrinsically defined periodicity. For our purposes, we mimic it by taking a one-dimensional SDE driven by a Brownian motion. It describes the motion of a state variable traveling through a bi-stable potential landscape, with an explicit periodic dependence of the potential shape creating a non-autonomous stochastic system that retains the characteristics of the two-dimensional model. Let $U$ be a double-well potential function, for example $U(T) = \frac{T^4}{4}-\frac{T^2}{2}, T\in \mathbb{R}.$ Moreover, fix a period length $T_0$, and some intensity parameter $c_1>0$. The diffusion process $T$ given by the SDE
$$dT_t = - U'(T_t)dt + c_1\cdot \sin\left(\frac{2\pi}{T_0} t \right) dt + \sigma dW_{t}$$
models temperature in a bi-stable environment. For $\sigma$ chosen appropriately, the trajectories of $K$ are almost periodic, with a tendency of alignment to the deterministic curves of period length $T_0$ given by the solution trajectories with noise turned off. This phenomenon is investigated under the name {\em stochastic resonance.} See \cite{herrmannimkellerpavljukevich03} for a review.\par\medskip

The second one comes from a 15-dimensional linear SDE of the Ornstein-Uhlenbeck type with a $15\times 15-$matrix with non-trivial rotational part and entries determined by satellite measurements which is used in linear prediction models for El Ni\~no (see \cite{penland96}). It creates a diffusion with non-trivial rotation numbers implying random periodicity for the sea surface temperature contained in the model. For our qualitative problems we may describe
the temperature curve as a simple mean-reverting linear sde with an additional deterministic periodic forcing. This leads to the following concrete example. A simple model for a temperature process fluctuating around an average value $A\in \mathbb{R}$ is given by an Ornstein-Uhlenbeck process determined by
\[ dT_t = \left[c_1 (T_t - A) + c_2 \sin \left( \frac{2 \pi}{T_0} t \right) \right]dt + \sigma dW_t, \]
where $c_1, c_2 > 0$, $\sigma > 0$ and $A\in\R$.
\end{ex}
In the next section we will povide sufficient conditions for the Regularity Properties 1 and 2 to be satisfied.
\subsection*{Regularity}
We will see that under some rather weak assumptions, the function $h^G$ is continuous on $[0,T] \times \R \times \R$ and belongs to the class $C^{1,2,2}([0,T[ \times \R \times \R) = C^{1,2}([0,T[ \times \R^2)$, and  $\psi^G$ is the solution of the linear PDE (\ref{linpde1}).
\begin{prop}[Regularity of the solution $h^G$] \label{regularity1}
Assume that:
\begin{enumerate}
\item (\ref{unifellip}) holds;
\item $b$, $\sigma$, $g$ are bounded functions;
\item $\sigma$ is Lipschitz continuous w.r.t. $x$;
\item $\psi^G(T,.) = \exp(-k G(.))$ is continuous and slowly increasing, i.e. for every $k > 0$:
$$\lim_{|x| \to +\infty} \exp(-kG(x)) \exp(-k|x|^2) = 0.$$
\end{enumerate}
Then the function $h^G$ is a classical solution, i.e. belongs to $C^{1,2}([0,T[ \times \R^{2};\R)$ and is continuous on $[0,T] \times \R \times \R$.
\end{prop}
\begin{proof}
In order to prove that $h^G$ is regular, we just have to prove that
$\psi^G$ is positive and smooth. But recall that $\psi = \psi^G$
solves the linear PDE (\ref{linpde1})
$$\frac{\partial \psi}{\partial t} + \frac{\sigma(t,x)^{2}}{2}
\frac{\partial^{2} \psi}{\partial x^{2}} + \left[ b(t,x) -
\frac{\be_{1} g(t,x) \sigma(t,x)}{\be^{2}} \right] \frac{\partial
\psi}{\partial x} - \frac{k g(t,x)^{2}}{2\be^{2} \eta} \psi = 0$$
with terminal condition:
$$\psi^G(T,x) = e^{kp} \exp ( -k G(x)).$$
Using the results of Veretennikov \cite{vereten80} and
\cite{vereten82}, we deduce that there exists a unique solution
$\psi^G$ in the space $\displaystyle \bigcap_{p > 1}
W^{1,2}_{p,loc}([0,T[ \times \R^2) \cap C([0,T] \times \R^2)$.
Moreover the solution $\psi^G$ is also slowly increasing.
\end{proof}

Proposition \ref{regularity1} implies that the Regularity Properties
1 and 2 are satisfied with $\mathcal{V} = \mathcal{U} = \R$; in the
first case $\psi^0(T,.) = 1$ is slowly increasing, while in the
second case $\psi^F(T,.) = \exp(-kF(.))$ is slowly increasing if $F$
is at most of linear growth:
$$\exists K > 0, \ |F(x)| \leq K(1+ |x|).$$
Notice that in the special case considered in \cite{muszar}, the
function is supposed to be bounded.
\begin{thm}\label{properties_nondeg}
Under the assumptions of the previous proposition, the
diversification pressure $\delta$, and therefore the optimal
strategies $\pi^{*}$ and $\pi^{\sharp}$, are locally bounded.
\end{thm}
\begin{proof}
Observe that the optimal strategies are given by
\begin{eqnarray} \label{optisharp}
\pi^{\sharp}(t,x) & = & \frac{g(t,x)}{\eta \be^{2}} + \frac{\be_{1}\sigma(t,x)}{k \be^{2}} \frac{1}{\psi^0(t,x)} \frac{\partial \psi^0}{\partial x} (t,x)
 \end{eqnarray}
and
\begin{eqnarray}\label{optistar}
\pi^{*}(t,x) & = & \frac{g(t,x)}{\eta \be^{2}} + \frac{\be_{1}\sigma(t,x)}{k \be^{2}} \frac{1}{\psi^F(t,x)} \frac{\partial \psi^F}{\partial x} (t,x).
\end{eqnarray}
The functions $g$ and $\sigma$ are bounded. Since $\psi^0(T,.) = 1$,
$\psi^0$ is uniformly bounded away from zero. Since $\exp (-kF) \geq
C
> 0$ on compact sets, $\psi^F$ stays locally away from zero.
Moreover, from \cite{vereten82}, we know that $\psi^G$ belongs to
$W^{1,2}_{p,loc}$ for all $p > 1$, thus $\nabla \psi^G$ is locally
bounded. Therefore the conclusion follows.
\end{proof}

\section{The geometric case}\label{case2}
%
%
In this section we assume that both the risk process $X$ and the
price process $P$ are geometric Brownian motions with drift. More
precisely, let $b(t,x) = \mu x$, $\sigma (t,x) = \nu x$ and $g =
\alpha$, where $\mu$, $\alpha \in \R$ and $\nu \in \R\setminus
\{0\}$. Having these rather strong assumptions on the diffusion, we
almost do not need to impose assumptions on the derivative function
$F$. All we need is a growth condition (see Equation
(\ref{intcond2})) which guarantees exponential integrability.
Neither do we need any continuity or smoothness conditions on $F$;
nor does $F$ have to be bounded, as it is assumed in \cite{muszar}.
 
\begin{ex}
Weather derivatives are financial instruments based on underlyings such as temperature, rainfall or snowfall.
A commonly used index are the accumulated heating degree days (cHDD).
A heating degree of a day with average temperature $T$ is defined as $HDD = \max\{0, 18 - T \}$. The cHDD are given as the sum of the HDD over a fixed period, for instance a month,
\[ \chdd = \sum_{i=1}^{31} \hdd_i. \]
The cHDD can be seen as a moving average process. Real data show that the cHDD are almost lognormally distributed and therefore they can be modeled as a geometric Brownian motion (see M. Davis \cite{davis}).
\end{ex}

Throughout we will assume that the initial value of $X$ satisfies
$X_0 = x_0 > 0$. The integral $\psi^G$ is given by
$$\psi^G(t,x) = \E \left[ \exp \left( -k G(x Z_{T-t}) \right) \exp \left( - \frac{\be_2^2}{2\be^{4}} \alpha^{2} (T-t) \right) \right],$$
where $Z_{r-t}$ is a Gaussian random variable with mean and variance
$$a = \left( \mu - \frac{\nu \al \be_{1}}{\be^{2}} - \frac{\nu^{2}}{2} \right)(r-t) \ \mbox{and} \ b^{2} = \nu^{2} (r-t).$$
First of all, the expectation in (\ref{solHJB}) must be finite. And a sufficient condition for this is that there is a constant $K$ such that
\begin{equation} \label{intcond2}
\forall x \geq 1, \ F(x) \geq - K(1+ \ln x).
\end{equation}
We continue by proving that this condition also implies Regularity Properties 1 and 2.

\subsection*{Regularity}
Notice that $h^0(t,x,v)= U(v)\exp \left( - \frac{\be_2^2}{2\be^{4}}
\alpha^{2} (T-t)\right)$, and Regularity Property 1 is trivially
satisfied. Regularity Property 2 will follow from the next theorem,
where the regularity of the function $h^F$ is directly verified. To
simplify notation we define
\[ I(t,x) = \E  \exp \left( -k F(x Z_{T-t}) \right), \]
which implies that we may write $h^F = \exp \left( - \frac{\be_2^2}{2\be^{4}} \alpha^{2} (T-t) \right)I(t,x)$.
\begin{prop}[Regularity of the solution $h^F$] \label{regularity2}
Assume (\ref{intcond2}). The function $h^F$ is a classical
solution, i.e. $h$ belongs to the class $C^{1,2}([0,T[ \times
\R^{*}_{+} \times \R;\R)$. Moreover if $F$ is continuous, $h^F$ is
in $C^{0}([0,T] \times \R_{+}^{*} \times \R;\R)$.
\end{prop}
\begin{proof}
We just have to prove that $I$ belongs to $C^{1,2}([0,T[ \times
\R^{*}_{+};\R)$. Observe that for all $x > 0$ and $t < T$
\begin{eqnarray*}
I(t,x) & = & \int_{\R} \exp \left( -k F(xe^{z}) \right)  \exp \left( - \frac{(z-a)^{2}}{2b^{2}} \right) \frac{dz}{b \sqrt{ 2\pi }} \\
& = & \int_{\R} \exp \left( -k F(e^{v}) \right)  \exp \left( - \frac{(v-\ln x -a)^{2}}{2b^{2}} \right) \frac{dv}{b \sqrt{ 2\pi }}.
\end{eqnarray*}
Therefore we deduce from the dominated convergence theorem, that
for all $x
> 0$ and $t < T$
\begin{eqnarray*}
\frac{\partial I}{\partial x} (t,x) & = & \int_{\R} \exp \left( -k F(e^{v}) \right)  \exp \left( - \frac{(v-\ln x -a)^{2}}{2b^{2}} \right)
\frac{(v- \ln x -a)}{b^{2}x} \frac{dv}{b \sqrt{ 2\pi }} \\
& = & \frac{1}{x} \int_{\R} \exp \left( -k F(xe^{z}) \right)  \exp \left( - \frac{(z
-a)^{2}}{2b^{2}} \right) \frac{(z -a)}{b^{2}} \frac{dz}{b \sqrt{ 2\pi }} \\
& = & \frac{1}{x} \E \left[ \exp \left( -k F(xe^{Z}) \right) \frac{(Z -a)}{b^{2}} \right].
\end{eqnarray*}
With similar arguments we obtain
\begin{equation*}
\frac{\partial^{2} I}{\partial x^{2}} (t,x) = \frac{1}{x^{2}} \E \left[ \exp \left( -k F(xe^{Z}) \right) \left( \frac{(Z -a)}{b^{2}} - \frac{1}{2} \right)^{2} \right] - \frac{4+b^{2}}{4b^{2}x^{2}} I(t,x).
\end{equation*}
Using once again the dominated convergence theorem, we prove that
\begin{eqnarray*}
\frac{\partial I}{\partial t} (t,x) & = & \frac{1}{2(T-t)} I(t,x) + \frac{a'}{b^{2}} \E \left[ \exp \left( -k F(xe^{Z}) \right) (Z -a) \right] \\
& & - \frac{1}{2 \nu^{2} (T-t)^{2}} \E \left[ \exp \left( -k F(xe^{Z}) \right) (Z -a)^{2} \right],
\end{eqnarray*}
with $a' = \left( \frac{\nu \al \be_{1}}{\be^{2}} + \frac{\nu^{2}}{2} - \mu \right)$.

If $F$ is continuous, then $I(T,x) = \exp (-kF(x))$ is also
continuous, and we deduce immediately that $h^F$ is continuous.
\end{proof}

%
\subsection*{Estimates of the diversification pressure}
Observe that the optimal strategy $\pi^{*}$, given
by (\ref{optimal1}), satisfies
\begin{equation*}
\pi^{*}(t,x) = \frac{\al}{\be^{2}} \left( 1 + \frac{\be_{1} \nu
x}{\al k} \frac{1}{I(t,x)} \frac{\partial I}{\partial x} (t,x)
\right),
\end{equation*}
and so the diversification pressure is given by
$$\displaystyle \delta(t,x) = \frac{\be_1 \nu}{k} \frac{x}{I(t,x)} \frac{\partial I}{\partial x}(t,x).$$
In the geometric case, the diversification pressure possesses the
following properties.
\begin{thm}\label{properties_geom}
If $F$ is bounded, then $\delta$ is finite on $[0,T[ \times
\R_{+}^{*}$. If $F$ is differentiable and satisfies
\begin{equation}\label{growth}
\exists M \geq 0, \ \forall x \geq 0, \ \left| x F'(x) \right|
\leq M, \end{equation} then
$$\forall t \in [0,T], \ \forall x > 0, \ \left| \delta (t,x) \right| \leq \be_1 \nu  M.$$\end{thm}

\begin{proof} Let $F$ be bounded. Recall (see Proposition
\ref{regularity2}) that for all $x
> 0$ and $t < T$
\begin{equation*}
\frac{\partial I}{\partial x} (t,x) = \frac{1}{b^{2}x} \E \left[
\exp \left( -k F(xe^{Z}) \right) (Z -a) \right].
\end{equation*}
Since $F$ is bounded, we deduce that
\begin{equation*}
\delta (t,x) = \frac{\be_1 \nu}{k} \frac{1}{b^{2}} \frac{\E \left[ \exp \left( -k F(xe^{Z}) \right) \left( Z-a \right) \right]}{\E \left[ \exp \left( -k
F(xe^{Z}) \right) \right]} \leq C \E \left[ \left| \frac{Z-a}{b^{2}} \right| \right] \leq \wtil{C} \frac{1}{b}.
\end{equation*}
The first result follows.

Now assume that $F$ is differentiable and that (\ref{growth})
holds. Since $I(T,.) = \exp (-k F(.))$, we have $\delta (T,x) = -
\be_1 \nu x F'(x)$, the absolute value of which is bounded by
$\be_1 \nu M$. Now for $t < T$ we have
\begin{equation*}
\frac{\partial I}{\partial x} (t,x) = (-k) \E \left[ e^{Z} F' \left( x e^{Z} \right) \exp \left( -k F(x e^{Z})\right) \right].
\end{equation*}
Hence
\begin{equation*}
\delta(t,x) = (-\be_1 \nu ) \frac{\E \left[ x e^{Z} F' \left( x e^{Z} \right) \exp \left( -k
F(x e^{Z})\right) \right]}{\E \left[ \exp \left( -k F(xe^{Z})\right) \right]}.
\end{equation*}
This implies
\begin{equation*}
\left| \delta(t,x) \right| \leq \be_1 \nu  M,
\end{equation*}
which achieves the proof.
\end{proof}

As an example for the second statement of the preceding Theorem, we
can take $F(x) = \ln (1+x)$ for $x \geq 0$. Recall that we always
assume that Condition (\ref{intcond2}) holds.

In the following example, Condition (\ref{growth}) does not hold.
Note that here, besides being differentiable, $F$ is non-decreasing.

\begin{ex}
If $F(x) = c (1+(\ln x)^{2})$ for $x \geq 0$ with a constant $c > 0$, then $\delta(t,x) = A_1(t) \ln x + A_2(t)$, where $A_1$ and $A_2$ are two positive bounded and continuous functions.
\end{ex}
\begin{proof}
We have
\begin{eqnarray*}
e^{kc} I(t,x) & = & \E \left[ \exp \left( -kc (\ln x + Z)^{2} \right) \right] \\
& = & e^{-kc (\ln x)^{2}} \E \left[ \exp \left( -2 kc \ln x Z - kc Z^{2} \right) \right] \\
& = & \frac{1}{\sqrt{1+2kcb^{2}}} \exp \left[ - \frac{kc}{1+2kcb^{2}} \left( (\ln x)^{2} - a \ln x + a^{2} \right) \right].
\end{eqnarray*}
Recall that
\begin{eqnarray*}
\frac{\partial I}{\partial x} (t,x) & = & \frac{1}{b^{2}x} \E \left[ \exp \left( -k F(xe^{Z}) \right) (Z - a) \right] \\
& = & -\frac{a}{b^{2}x}I(t,x) + \frac{1}{b^{2}x} \E \left[ \exp \left( -k F(xe^{Z}) \right) Z \right],
\end{eqnarray*}
so that
$$\delta(t,x)  =  -\be_1 \nu \frac{a}{kb^{2}} + \be_1 \nu \frac{1}{kb^{2}I(t,x)} \E \left[ \exp \left( -k F(xe^{Z}) \right) Z \right].$$
Now
$$e^{kc} \E \left[ \exp \left( -k F(xe^{Z}) \right) Z \right] = \frac{a - 2kcb^{2} \ln x}{1+2kcb^{2}} I(t,x),$$
which leads to
$$\delta(t,x)  = -\be_1 \nu  \frac{2c}{1+2kcb^{2}} (a + \ln x).$$
\end{proof}

We remark that the bound on the functions $A_1$ and $A_2$ depends
linearly on $c$. If $c$ is sufficiently small, then we can apply
Lemma \ref{quasiadmi}.


The situation of the preceding example can be generalized to give the following result.

\begin{thm}
If $F$ is Lipschitz and non-decreasing, then there is a constant $K$ such that
$$\left| \delta(t,x) \right| \leq K(1+\left| x \right|)$$
and $\delta$ is non-increasing.
\end{thm}

\begin{proof}
Recall that
$$\frac{\partial I}{\partial x}(t,x)  = (-k) \E \left[ \exp \left( -k F(x e^{Z}) \right) e^{Z} \frac{\partial F}{\partial x}(x
e^{z})\right].$$
Suppose that $\frac{\partial F}{\partial x}$ is bounded by $M$. Then
\begin{eqnarray*}
\left| \frac{\partial I}{\partial x}(t,x) \right| & \leq & k M \E \left[ \exp \left( -k F(x e^{Z}) \right) e^{Z} \right] \\
& = & k M e^{a+b^{2}/2} \E \left[ \exp \left( -k F(x e^{b^{2}} e^{b N+a}) \right) \right],
\end{eqnarray*}
where $N$ is a standard Gaussian r.v. Hence
$$\left| \frac{1}{I(t,x)} \frac{\partial I}{\partial x}(t,x) \right| \leq k M e^{a+b^{2}/2} \frac{\E \left[ \exp \left( -k F(x
e^{b^{2}} e^{bN+a}) \right) \right]}{\E \left[ \exp \left( -k F(x
e^{bN+a}) \right) \right]}.$$ If $F$ is increasing, then for all
$(t,x) \in [0,T[ \times \R^{*}_{+}$ we have
\begin{equation} \label{strategyestimate}
 \left| \frac{1}{I(t,x)} \frac{\partial I}{\partial x}(t,x) \right| \leq k M e^{a+b^{2}/2},
\end{equation}
and
$$\frac{1}{I(T,x)} \frac{\partial I}{\partial x}(T,x) = (-k) \frac{\partial F}{\partial x}(x).$$
So Inequality (\ref{strategyestimate}) holds for $t=T$.
\end{proof}
\section{Admissibility of the optimal cross hedging strategy} \label{secadmi}

We will now analyze under which conditions the optimal strategies
$\pi^\sharp$ and $\pi^*$ are quasi-admissible. To this end we will
frequently use the following {\em exponential inequality}. If $M$
is a continuous local martingale vanishing at $0$, then
\begin{equation} \label{170805-1}
\prb \left[ \sup_{t \ge 0} |M_t| \ge x, \langle M, M \rangle_\infty  \le y \right] \le \exp \left(\frac{-x^2}{2y}\right),
\end{equation}
(see Chapter IV in \cite{RY}).

In order to find sufficient conditions for the optimal strategies
$\pi^\sharp$ and $\pi^*$ to be quasi-admissible we first provide
conditions under which the derivatives of $\psi^0$ and $\psi^F$
are bounded.
\begin{lem}\label{boundedpsi}
Suppose that there exists a constant $C$ such that $\sigma$ and
$\widehat b$ are globally Lipschitz continuous and for all $(t,x)
\in [0,T] \times \R$ we have $\frac{|\widehat b(t,x)| +
|\sigma(t,x)|}{1 + |x|} \le C.$

If $g$ is bounded and Lipschitz continuous in $x$, then the
partial derivative $\frac{\partial \psi^0}{\partial x}$ is
bounded; and if in addition $F$ is Lipschitz continuous and bounded from below, then
$\frac{\partial \psi^F}{\partial x}$ is bounded. 
\end{lem}
\begin{proof}
We first show boundedness of $\frac{\partial \psi^F}{\partial x}$.
Observe that there exist constants $C_1, C_2, \ldots$ such that
for all $x$, $x' \in \R$ and $t \in [0,T]$,
\begin{eqnarray*}
&& |\psi^F(t,x) - \psi^F(t,x')| \\
&\le& C_1 \Big\{ \E\left[\exp(-kF(Y^{t,x}_T)) \Big|\exp(-\frac{\be_2^2}{2\beta^4} \int_t^T g^2(r, Y^{t,x}_r)dr) - \exp(-\frac{\be_2^2}{2\beta^4}\int_t^T g^2(r, Y^{t,x'}_r)dr)\Big| \right] \\
&&+ \ \E\left[\exp\left(-\frac{\be_2^2}{2\beta^4} \int_t^T g^2(r, Y^{t,x'}_r)dr\right) |\exp(-kF(Y^{t,x}_T)) - \exp(-kF(Y^{t,x'}_T))| \right] \Big\}\\
&\le& C_2 \Big\{ \E\left[ \sup_{r\in[t,T]}|g^2(r, Y^{t,x}_r) - g^2(r, Y^{t,x'}_r)| \right] + \E\left[|\exp(-kF(Y^{t,x}_T)) - \exp(-kF(Y^{t,x'}_T))| \right] \Big\}\\
&\le& C_3 \Big\{ \E\left[\sup_{r\in[t,T]}|g(r, Y^{t,x}_r) - g(r, Y^{t,x'}_r)| \right] + \E\left[|F(Y^{t,x}_T) - F(Y^{t,x'}_T)| \right] \Big\} \\
&\le& C_4 \E\left[\sup_{r\in[t,T]} |Y^{t,x}_r-Y^{t,x'}_r| \right].
\end{eqnarray*}
It follows from standard results (see Lemma 4.5.6. in
\cite{kunita}) that $\E\left[\sup_{r\in[t,T]}
|Y^{t,x}_r-Y^{t,x'}_r| \right] \le C_5 |x - x'|$. Consequently all
difference quotients $\frac{\psi^F(t,x) - \psi^F(t,x')}{x-x'}$ are
uniformly bounded, and hence $\frac{\partial \psi^F}{\partial x}$
is bounded. Similarly, one can show that $\frac{\partial
\psi^0}{\partial x}$ is bounded as well.
\end{proof}

\begin{corol}\label{corobounded}
Let the assumptions of Lemma \ref{boundedpsi} be satisfied and suppose that $\sigma$ is bounded. Then the optimal strategies $\pi^\sharp$ and $\pi^*$ are bounded.
\end{corol}
\begin{proof}
This follows immediately from (\ref{optisharp}) and (\ref{optistar}).
\end{proof}
\begin{thm}\label{quasiadmi_drift1}
Let the assumptions of Corollary \ref{corobounded} be satisfied. Then $U(V^\sharp_T)$ and $U(V^*_T)$ are integrable; and $\pi^\sharp$ and $\pi^*$ are quasi-admissible.
\end{thm}
\begin{proof}
Let $\pi$ be either $\pi^\sharp$ or $\pi^*$. By Corollary \ref{corobounded}, $\pi$ is bounded, and hence $U(V^{\pi}_t)$ belongs to $L^{1}$ for all $t\in[0,T]$.

For $n\in\mathbb{N}$ let $\tau_{n}$ be the following stopping
time:
$$\tau_{n} = \inf \left\{t \geq 0, \ V^{\pi}_{t} = -n \right\} \wedge T.$$
The process $V^{\pi}_{t \wedge \tau_{n}}$ is bounded from below by $-n$, and we have
\begin{eqnarray*}
&& \E |U(V^{\pi}_{T}) - U(V^{\pi}_{\tau_{n}})|  =  \E \left[ |\exp (-\eta V^{\pi}_{T}) - \exp (\eta n)| \ind_{\{\tau_{n} < T\}} \right] \\
&& \quad = \E \left[ \left( \exp (-\eta V^{\pi}_{T}) - e^{\eta n} \right) \ind_{\tau_{n} < T} \ind_{\{V^{\pi}_{T} \leq -n\}} \right] \\
&& \quad \quad + \E \left[ \left( e^{\eta n} - \exp (- \eta V^{\pi}_{T}) \right) \ind_{\{\tau_{n} < T\}} \ind_{\{V^{\pi}_{T} >  -n\}} \right] \\
&& \quad \leq \E \left[ \exp (-\eta V^{\pi}_{T})
\ind_{\{V^{\pi}_{T} \leq -n\}} \right] + e^{\eta n} \prb (\tau_{n}
< T).
\end{eqnarray*}
Notice that $P(\tau_{n} < T) \le P(\sup_{t \in [0,T]} |V^\pi_t|
\ge n)$. Obviously the quadratic variation of $V^\pi_t$ is
bounded, say by $B \in \R_+$. Hence the exponential inequality
(\ref{170805-1}) yields
\[ P(\tau_{n} < T) \le \exp\left( \frac{-n^2}{2B}\right). \]
As a consequence we have $\displaystyle \lim_{n \to \infty}  \E |U(V^{\pi}_{T}) - U(V^{\pi}_{\tau_{n}})| = 0$, and hence the result.
\end{proof}

In the remainder of this section we consider some special cases of our model. We can thus weaken the assumptions of Theorem \ref{quasiadmi_drift1}.

\subsection*{The case $\beta_1 = 0$}
Suppose in the following that $\beta_1=0$, which means that the
external risk and the price process are only correlated via the
drift part. Notice that in this case $\pi^\sharp_t = \pi^*_t =
\frac{g(t,X_{t})}{\eta \be_{2}}$, and in particular the optimal
strategy $\pi^*$ does not depend on the structure of the
derivative $F(X_T)$. We also assume that $g$ is such that the
stochastic integral of $g(\cdot,X)$ relative to $B$ is defined.
This is guaranteed for instance if $g$ is bounded or continuous in
$t$ and $x$.
\begin{thm}\label{quasiadmi_drift}
Let $\beta_1 = 0$. Then $U(V^*_T)$ is integrable and $\pi^*$ is quasi-admissible.
\end{thm}
\begin{proof}
We put $M_t = \int_0^t \frac{g(u,X_{u})}{\be_{2}} dB_{u}$. Then $V^*_t = v_0 + \frac1\eta M_t + \frac1\eta \langle M,M \rangle_t$, and
\[ \E e^{-\eta V^*_T} \le e^{-\eta v_0} \E e^{- M_T - \frac{1}{2} \langle M,M \rangle_T} \le e^{- \eta v_0}, \]
which means that $U(V^*_T)$ is integrable.

Let $\tau_n = \inf\{t\ge 0: V^*_ t = -n \} \wedge T$,
$n\in\mathbb{N}$. Then the strategies $\pi^n = \pi^* 1_{[0,
\tau_n]}$ are admissible, $V^{\pi_n}_T = V^*_{\tau_n}$, and
\begin{eqnarray*}
\E |U(V^*_{T}) - U(V^*_{\tau_{n}})|  &=&  \E \left[ |\exp (-\eta V^{*}_{T}) - \exp (\eta n)| \ind_{\{\tau_{n} < T\}} \right] \\
&\leq& \E \left[ \exp (- \eta V^*_{T}) \ind_{\{V^{*}_{T} \leq
-n\}} \right] + e^{\eta n} \prb (\tau_{n} < T).
\end{eqnarray*}
The maximal inequality for positive supermartingales (see f.e.
\cite{RY}, Chapter II), and Inequality (\ref{170805-1}) imply
\begin{eqnarray*}
\prb(\tau_{n} < T) &=& \prb(\sup_{t \in [0, T]} [-M_t - \langle M,M \rangle_t] \ge \eta n + v_0) \\
&\le& \prb(\sup_{t \in [0, T]}[-M_t - \frac12 \langle M,M
\rangle_t] \ge \frac98
\eta n + v_0, \langle M,M \rangle_T \ge \frac{\eta n}{4}) \\
& & + \ \prb(\sup_{t \in [0, T]}  [-M_t - \langle M,M \rangle_t] \ge \eta n + v_0, \langle M,M \rangle_T \le \frac{\eta n}{4}) \\
&\le& \prb(\sup_{t \in [0, T]} e^{-M_t - \frac12 \langle M,M \rangle_t} \ge e^{\frac98 \eta n + v_0}) \\
&& + \prb(\sup_{t \in [0, T]} -M_t \ge \eta n+ v_0, \langle M,M \rangle_T \le \frac{\eta n}{4}) \\
&\le& e^{-\frac98 \eta n - v_0} + e^{-2\frac{(\eta n+ v_0)^2}{\eta n}}.
\end{eqnarray*}
Therefore $\lim_{n\to \infty} e^{\eta n} \prb (\tau_{n} < T) = 0$.
\end{proof}

\subsection*{The geometric case} \label{admissibility}

As in Section \ref{case2} we consider here the case that the risk
process $X$ and the price $P$ are geometric Brownian motions. To
this end, assume again $b(t,x) = \mu x$, $\sigma (t,x) = \nu x$
and $g = \alpha \in \R$. First observe that in this case
$\pi^\sharp = \frac{\alpha}{\eta \beta^2}$.
\begin{thm} \label{040805-1}
The optimal strategy $\pi^\sharp$ maximizing the right-hand side
of (\ref{price}) is quasi-admissible.
\end{thm}
\begin{proof}
We have already proved that $U(V^{\sharp})$ belongs to $L^{1}$.
For $n\in\mathbb{N}$ let $\tau_{n}$ be the following stopping
time:
$$\tau_{n} = \inf \left\{t \geq 0, \ V^{\sharp}_{t} = -n \right\} \wedge T.$$
The process $V^{\sharp}_{t \wedge \tau_{n}}$ is bounded from below by $-n$, and we have:
\begin{eqnarray*}
&& \E |U(V^{\sharp}_{T}) - U(V^{\sharp}_{\tau_{n}})|  =  \E \left[ |\exp (-\eta V^{\sharp}_{T}) - \exp (\eta n)| \ind_{\{\tau_{n} < T\}} \right] \\
&& \quad = \E \left[ \left( \exp (-\eta V^{\sharp}_{T}) - e^{\eta n} \right) \ind_{\{\tau_{n} < T\}} \ind_{\{V^{\sharp}_{T} \leq -n\}} \right] \\
&& \quad \quad + \E \left[ \left( e^{\eta n} - \exp (- \eta V^{\sharp}_{T}) \right) \ind_{\{\tau_{n} < T\}} \ind_{\{V^{\sharp}_{T} >  -n\}} \right] \\
&& \quad \leq \E \left[ \exp (-\eta V^{\sharp}_{T}) \ind_{V^{\sharp}_{T} \leq -n} \right] + e^{\eta n} \prb (\tau_{n} < T).
\end{eqnarray*}
If $\tau_{n} < T$, then $\displaystyle \sup_{ s\in[0,T]}
|\wtil{W}_{s} | \geq \eta \frac{n+v_{0}}{\al}$, with $\displaystyle
\wtil{W}_{s} = \frac{\be_{1}}{\be^{2}} W_{s} +
\frac{\be_{2}}{\be^{2}} B_{s}$. Thus,
$$e^{\eta n} \prb (\tau_{n} < T) \leq  e^{\eta n} \exp \left( - \frac{\eta^2
(n+v_{0})^{2}}{2\al^{2}T} \right),$$
and hence $\displaystyle \lim_{n \to \infty}  \E |U(V^{\sharp}_{T}) - U(V^{\sharp}_{\tau_{n}})| = 0$.
\end{proof}

We now give a sufficient criterion for the optimal strategy
$\pi^*$ to be quasi-admissible. This criterion was seen to be
valid in important scenarios in Section \ref{case2}.
\begin{thm} \label{quasiadmi}
Suppose that there exist constants $A, C > 0$ such that the diversification pressure satisfies $|\delta(u, x)| \le C + A |\log x|$, for all $(u, x) \in [0, T] \times \R^*_+$. If
$A$ is sufficiently small, then the optimal control $\pi^*$ is quasi-admissible.
\end{thm}
\begin{proof}
To simplify notation we assume throughout that the constant $C$
may be chosen to be zero, i.e.\ that $|\delta(u, x)| \le A |\log
x|$ for $(u, x) \in [0, T] \times \R^*_+$. For $n\in\mathbb{N}$
let $\tau_n = \inf\{t\ge 0: V^*_ t = -n \} \wedge T$. Then the
strategies $\pi^n = \pi^* 1_{[0, \tau_n]}$ are admissible, we have
$V^{\pi_n}_T = V^*_{\tau_n}$, and
\begin{eqnarray*}
\E |U(V^*_{T}) - U(V^*_{\tau_{n}})|  &=&  \E \left[ |\exp (- \eta V^{*}_{T}) - \exp (\eta n)| \ind_{\{\tau_{n} < T\}} \right] \\
&\leq& \E \left[ \exp (- \eta V^*_{T}) \ind_{\{V^{*}_{T} \leq
-n\}} \right] + e^{\eta n} \prb (\tau_{n} < T).
\end{eqnarray*}
Note that by Equation (\ref{020905-1}) we have $V^*_t = V^{\sharp}_{t} + \int_0^t \kappa \delta(u, X_u) \frac{1}{P_u} \, dP_u$, where $\kappa = \frac{1}{\beta^2}$.
Therefore
\begin{eqnarray*}
\prb(\tau_n < T) &=& \prb(\inf_{t \in [0, T]} V^*_t \le -n) \\
&\le& \prb(\inf_{t \in [0, T]} V^\sharp_t \le -\frac{n}{3}) + \prb(\inf_{t \in [0, T]} \int_0^t \kappa \delta(u, X_u) \alpha du \le -\frac{n}{3}) \\
& &  + \ \prb(\inf_{t \in [0, T]} \int_0^t \kappa \delta(u, X_u) (\beta_1 dW_u + \beta_2 dB_u) \le -\frac{n}{3}).
\end{eqnarray*}
By arguments as in the proof of Lemma \ref{040805-1}
$$D_1(n) = \prb(\inf_{t \in [0, T]} V^\sharp_t \le -\frac{n}{3}) \le \exp\left( -
\frac{\eta^2 (\frac{n}{3} + v_0)^2}{2 \alpha^2 T} \right).$$
Furthermore
\begin{eqnarray*}
D_2(n) &=& \prb\left(\inf_{t \in [0, T]} \int_0^t \kappa \delta(u,X_u) \alpha du \le -\frac{n}{3} \right) \\
&\le& \prb\left(\sup_{t \in [0, T]} \int_0^t A |\kappa \alpha| |\log{X_u}| du \ge \frac{n}{3}\right) \\
&\le& \prb\left( \int_0^T A |\kappa \alpha| |\nu W_u + (\mu - \frac{\nu^2}{2})u| du \ge \frac{n}{3}\right) \\
&\le& \prb\left(\sup_{t \in [0, T]} |W_u| \ge \frac{ \frac{n}{3 A |\kappa \alpha|} - |\mu-\frac{\nu^2}{2}| \frac{T^2}{2}}{\nu T}\right) \\
&\le& \exp\left( - \frac{\left( \frac{n}{3 A |\kappa \alpha| \nu T} - |\mu-\frac{\nu^2}{2}|\frac{T}{2\nu}\right)^2}{2T} \right).
\end{eqnarray*}
Finally, the third summand satisfies
\begin{eqnarray*}
&& D_3(n) = \prb( \inf_{t \in [0, T]} \int_0^t \kappa \delta(u, X_u) (\beta_1 dW_u + \beta_2 dB_u) \le -\frac{n}{3}) \\
&& = \prb \left( \inf_{t \in [0, T]} \int_0^t \delta(u, X_u) (\beta_1 dW_u + \beta_2 dB_u) \le -\frac{n}{3\kappa }, \right. \\
&& \qquad \qquad \qquad \left. \mbox{and} \ \int_0^T \delta(u, X_u)^2 du \le \frac{A n}{\beta^2} \right) \\
&& + \prb \left(\inf_{t \in [0, T]} \int_0^t \delta(u, X_u) (\beta_1 dW_u + \beta_2 dB_u) \le -\frac{n}{3\kappa}, \right. \\
&& \qquad \qquad \qquad \left. \mbox{and} \  \int_0^T \delta(u, X_u)^2 du > \frac{A n}{\beta^2} \right) \\
&&\le \exp\left( - \frac{\frac{n^2}{9 \kappa^2}}{2 n A} \right) + \prb\left(\int_0^T A^2 \log (X_u)^2 \beta^2 du > A n \right) \\
&&\le \exp\left( - \frac{n}{18 A \kappa^2} \right) + \prb\left(
\int_0^T |\nu W_u + \mu u - \frac{\nu^2}{2} u|^2 du > \frac{n}{A
\beta^2} \right).
\end{eqnarray*}
Moreover
\begin{eqnarray*}
&& \prb \left( \int_0^T |\nu W_u + \mu u - \frac{\nu^2}{2} u|^2 du > \frac{n}{A \beta^2} \right) \\
&& \qquad \le \prb \left( \int_0^T 2\nu^2 W_u^2 + 2(\mu-\frac{\nu^2}{2})^2 u^2 du > \frac{n}{A \beta^2} \right) \\
&& \qquad \le \prb \left( 2\nu^2 T \sup_{t \in [0, T]} |W_u|^2 >  \frac{n}{A \beta^2} - \frac23 (\mu-\frac{\nu^2}{2})^2 T^3 \right) \\
&& \qquad \le \exp\left( - \frac{ \left| \frac{n}{2 A \beta^2 \nu^2 T} - \frac{(\mu-\frac{\nu^2}{2})^2 T^2}{3 \nu^2} \right| } {2T} \right)
\\
&& \qquad = \exp\left( - \left| \frac{n}{4 A \beta^2 \nu^2 T^2} - \frac{(\mu-\frac{\nu^2}{2})^2 T}{6 \nu^2} \right|\right).
\end{eqnarray*}
As a consequence, if $A$ is sufficiently small, then
$$\lim_{n \to \infty} e^{\eta n} \prb (\tau_{n} < T) = \lim_{n \to \infty} e^{\eta n} (D_1(n) + D_2(n) + D_3(n)) = 0,$$
and thus the proof is complete.
\end{proof}
\section{Dynamic versus static risk}\label{dynamic_static}
%
Consider an insurer who sells a derivative. The contract is such
that at maturity $T$ he has a risk source related income given by
$F(X_T)$. We may view the contract from his perspective as
compelling him to pay $-F(X_T)$ at maturity. Hence the expected
value $\E U(F(X_T))$ measures the static risk of making this
payment at time $T$. By how much does dynamic cross hedging reduce
this risk?

Let $s$ be the insurer's indifference price of the derivative $F(X_T)$ if he does {\em not} invest in a correlated financial asset. $s$ is uniquely determined by the equation $\E U(F(X_T) - s) = \E U(0) = -1$, and we refer to $s$ as the {\em static indifference price} of $F(X_T)$. Obviously $s$ satisfies
\[ s = - \frac{1}{\eta} \log \E e^{- \eta F(X_T)}. \]
Note that the static indifference price is again independent of the initial wealth.

The indifference prices $p$ and $s$ measure the risk associated with the insurance derivative $F(X_T)$. Indeed $p$ and $s$ coincide with a so-called {\em entropic risk}. We recall the definition of this concept.
\begin{defin}
We denote by $\X$ the set of all measurable random
variables. The {\em entropic risk measure} $\rho_\gamma$ with {\em
risk aversion coefficient} $\gamma > 0$ is defined by
\[ \rho_\gamma: \X \ni \Psi \mapsto \frac{1}{\gamma} \log \E e^{-\gamma \Psi}. \]
We will write $\rho_\gamma^\prb$ if we want to stress the probability measure $\prb$ we are referring to.

It can be shown that the mapping
$$m_\gamma: \X \ni \Psi \mapsto \inf \{ \rho_\gamma(\Psi - V^\pi_T + v_0): \pi \textrm{ is a strategy}\}$$
is also a risk measure for any $\gamma > 0$ (see Barrieu, El Karoui \cite{bar}). It measures the risk by taking into account the cross hedging strategies with initial capital $v_0 = 0$. We call $m_\gamma$ therefore the {\em cross modified risk measure}.
\end{defin}

Observe that the static indifference price of a derivative $F(X_t)$ coincides with the negative entropic risk of our insurance derivative, i.e.
\[ s = - \rho_\eta(F(X_T)). \]
Remarkably, the indifference price $p$ can also be viewed as entropic risk. Note that by Equation (\ref{ptexpression}),
\begin{equation} \label{newrisk}
p = - \rho_k^{\qprb} (F(Y_T^{0,x_0})),
\end{equation}
where $\qprb$ is the measure defined in Equation (\ref{newq}) and $Y^{0,x}$ is the solution of the SDE (\ref{crossSDE}) with cross modified drift $\displaystyle \widehat{b}(t,x) = b(t,x) - \frac{\be_{1} g(t,x) \sigma(t,x) }{\be^{2}}$. The dynamic indifference price is thus the entropic risk of the cross modified payoff $F(Y_T)$ relative to the {\em new} probability measure $\qprb$ and the risk aversion coefficient $k$.

If $\int \frac{\be_1}{\be^2}g(\cdot,X^{\cdot,x_0}) dW$ satisfies
Novikov's condition, then by Theorem \ref{priceunterqhat} we have
\begin{equation} \label{pricenovi2}
p = - \rho_k^{\widehat \qprb} (F(X_T^{0,x_0})) ,
\end{equation}
where $\widehat \qprb$ is defined as in (\ref{newqhat}).
Consequently, we have the following result:
\begin{thm}\label{090805-1}
Let $\ID$ be the set of all derivatives of the form $F(X_T)$. If $\int \frac{\be_1}{\be^2}g(\cdot,X^{\cdot,x_0}) dW$ satisfies
Novikov's condition, then the cross modified risk measure
$m_{\eta}$ on $\ID$ is equal to the entropic risk measure relative
to the new probability measure $\widehat \qprb$ and risk
coefficient $k$, i.e.\
\[ m_{\eta} = \rho_k^{\widehat \qprb} \quad \textrm{ on }\ID. \]
\end{thm}
We thus are able to to compare the risk associated with $F(X_T)$
if we invest and if we do not invest into the correlated asset:
the difference $p-s$ is a {\em monetary} measure of the risk
reduction through dynamic hedging.

Observe that, since $k < \eta$, cross hedging always leads to a
reduction of the risk aversion coefficient.  Surprisingly however,
the dynamic risk is {\em not necessarily} smaller than the static
risk. This may be the case if the derivative $F(X_T)$ amplifies
the risk of the portfolio and the negative effect due to lower
diversification exceeds the positive effect due to the aversion
reduction (see Proposition \ref{drift}).

If the derivative $F(X_T)$ diversifies the risk in our portfolio,
then the dynamic risk will always be smaller than the static risk.
For example, if $F$ is monotone increasing, $g \ge 0$ and
$\frac{\be_{1} \sigma }{\be^{2}} \le 0$, then the derivative has a
diversifying impact. It does not come as a surprise in this case
that the risk is always reduced by cross hedging, as is shown in
the next result.
\begin{prop}\label{riskred}
Let $F$ be monotonically increasing, $g \ge 0$ and $\frac{\be_{1}
\sigma }{\be^{2}} \le 0$ and suppose that
\[ \mathrm{cov} \left(\exp(- k F(Y^{0,x_0}_T)), \exp \left[ - \frac{k}{2 \eta \be_{2}^{2}} \int_{0}^{T} g^2(r,Y^{0,x_0}_{r})dr \right] \right) \le 0. \]
Then dynamic hedging reduces the risk of $F(X_T)$; in other words, $p-s \ge 0$.
\end{prop}
\begin{proof}
Let $Y= Y^{0,x}$ be the solution of the SDE (\ref{crossSDE}), and define the abbreviations
$$\zeta = \exp(-\eta F(X_T)), \ \Psi= \exp(- k F(Y_T)) \ \mbox{and} \ \xi = \exp \left[ - \frac{k}{2 \eta \be_{2}^{2}} \int_{0}^{T} g^2(r,Y_{r})dr \right].$$
The drift of $Y_t$, given by $\displaystyle \widehat{b}(t,x) =
b(t,x) - \frac{\be_{1} g(t,x) \sigma(t,x) }{\be^{2}}$, is bigger
than the drift $b(t,x)$ of the SDE $X$. If we choose $\widehat W =
W$ in (\ref{crossSDE}), then by a comparison theorem for SDEs (see
Theorem 3.7, Ch. IX, \cite{RY}), we have $X_T \le Y_T$, $P$-a.s.
Then $\zeta = e^{-k F(X_T)} \ge e^{-k F(Y_T)}$, and with Jensen's
Inequality, $\E(\zeta)^{\frac{k}{\eta}} \ge \E(\Psi)$. Therefore,
\begin{equation*}
p - s = -\log \left( \frac{  \E \left( \Psi \xi \right) }{ \E \left(\zeta \right)^{\frac{k}{\eta}}  \E \left( \xi \right) }\right)^{\frac1k}
\ge -\log \left( \frac{  \E \left( \Psi \xi \right) }{ \E(\Psi) \E \left( \xi \right) }\right)^{\frac1k} = -\log \left(1
 +\frac{\mathrm{cov} \left(\Psi, \xi \right)}{\E(\Psi) \E(\xi)} \right),
\end{equation*}
and thus, $p- s \ge 0$ if $\mathrm{cov} \left(\Psi, \xi \right) \le 0$.
\end{proof}

%
%
%
Consider now the special case where the price process $P$ is only
driven by the second Brownian motion $B$, i.e. $\beta_1 = 0$. Then
$P$ and the risk process $X$ are only correlated via the drift
term $g$. In this case $k = \eta$, and hence cross hedging does
not reduce the aversion coefficient. Moreover, the negative
correlation between the drift $g$ and the derivative $F(X_T)$ is a
sufficient and {\em necessary} condition for cross hedging to
reduce risk.
\begin{prop}\label{drift}
Let $\beta_1 = 0$. Then $p \ge s$ if and only if
$$\mathrm{cov} \left( \exp(-\eta F(X_T)), \exp \left[ - \frac{k}{2 \eta \be_{2}^{2}} \int_{0}^{T} g^2(r,X_{r})dr \right] \right) \le 0.$$
\end{prop}
\begin{proof}
Note that the random variables $\zeta$ and $\Psi$ defined in the proof of Proposition \ref{riskred} coincide if $\beta_1 = 0$. Therefore
\begin{eqnarray*}
 p - s = -\log \left(1  +\frac{\mathrm{cov} \left(\zeta, \xi \right)}{\E \left( \zeta \xi \right)} \right),
\end{eqnarray*}
which implies the result.
\end{proof}

Let us finally consider the case where the drift of the price
process satisfies $g = 0$. Normally a risk averse investor would
not invest in an asset with price dynamics $P$. However, if he has
derivative $F(X_T)$ in his portefolio, then he should invest
$\frac{1}{\beta^2} \delta(t, X_t)$ of his wealth at time $t$ in
the asset. Doing this he will always reduce his risk.
\begin{prop} \label{riskcomparison}
Let $g = 0$. Then the dynamic indifference price $p$ of a
derivative $F(X_T)$ is always bigger than the static price $s$.
\end{prop}
\begin{proof}
Note that in this case
\begin{equation} \label{080805-1}
p - s = \log \left( \frac{\E[e^{-\eta F(X_T)}]^{\frac{1}{\eta} }}{\E[e^{-k F(X_T)}]^\frac1k }\right).
\end{equation}
Since $k < \eta$, Jensen's inequality implies $\E[e^{-k F(X_T)}]^\frac1k \le \E[e^{-\eta F(X_T)}]^\frac{1}{\eta}$, and thus $p - s \ge 0$.
\end{proof}

\bibliography{control2}

\providecommand{\bysame}{\leavevmode\hbox to3em{\hrulefill}\thinspace}
\providecommand{\MR}{\relax\ifhmode\unskip\space\fi MR }
\providecommand{\MRhref}[2]{%
  \href{http://www.ams.org/mathscinet-getitem?mr=#1}{#2}
}
\providecommand{\href}[2]{#2}
\begin{thebibliography}{10}

\bibitem{fangbarcilonwang99}
A.~Barcilon, Z.~Fang, and B.~Wang, \emph{Stochastic dynamics of {El
  Nino}-southern oscillation}, J. Atmos. Sci. \textbf{856} (1999), 5--23.

\bibitem{bar}
P.~Barrieu and N.~El~Karoui, \emph{Optimal risk transfer}, 2004.

\bibitem{chaumontimkellermueller05}
S.~Chaumont, P.~Imkeller, and M.~M\"uller, \emph{Equilibrium trading of climate
  and weather risk and numerical simulation in a markovian framework}, SERRA
  (2005, to appear).

\bibitem{davis2}
M.~Davis, \emph{Optimal hedging with basis risk}, Preprint.

\bibitem{davis}
\bysame, \emph{Pricing weather derivatives by marginal value}, Quant. Finance
  \textbf{1} (2001), no.~3, 305--308.

\bibitem{FS}
W.H. Fleming and H.M. Soner, \emph{Controlled markov processes and viscosity
  solutions.}, Applications of Mathematics. 25. New York: Springer-Verlag. xv,
  428 p., 1993.

\bibitem{herrmannimkellerpavljukevich03}
S.~Herrmann, P.~Imkeller, and I.~Pavlyukevich, \emph{Two mathematical
  approaches of stochastic resonance}, Interacting Stochastic Systems (J.-D.
  Deuschel and A.~Greven, eds.), Springer, 2004, pp.~327--351.

\bibitem{huimkellermueller04}
Y.~Hu, P.~Imkeller, and M.~M\"uller, \emph{Partial equilibrium and market
  completion}, Int. J. Theor. Appl. Finance \textbf{8} (2005), 483--508.

\bibitem{kunita}
H.~Kunita, \emph{Stochastic flows and stochastic differential equations},
  Cambridge Studies in Advanced Mathematics, vol.~24, Cambridge University
  Press, Cambridge, 1990.

\bibitem{monoyios}
M.~Monoyios, \emph{Performance of utility-based strategies for hedging basis
  risk}, Quant. Finance \textbf{4} (2004), no.~3, 245--255.

\bibitem{muszar}
M.~Musiela and T.~Zariphopoulou, \emph{{An example of indifference prices under
  exponential preferences.}}, Finance Stoch. \textbf{8} (2004), no.~2, 229--239
  (English).

\bibitem{penland96}
C.~Penland, \emph{A stochastic model of indo pacific sea surface temperature
  anomalies}, Physica D \textbf{98} (1996), 534--558.

\bibitem{RY}
D.~Revuz and M.~Yor, \emph{Continuous martingales and {B}rownian motion},
  Grundlehren Math. Wiss., vol. 293, Springer-Verlag, Berlin Heidelberg New
  York, 1991.

\bibitem{elkarouirouge}
R.~Rouge and N.~El~Karoui, \emph{Pricing via utility maximization and entropy},
  Math. Finance \textbf{10} (2000), no.~2, 259--276, INFORMS Applied
  Probability Conference (Ulm, 1999). \MR{MR1802922 (2001m:91066)}

\bibitem{schach01}
W.~Schachermayer, \emph{Optimal investment in incomplete markets when wealth
  may become negative}, Ann. Appl. Probab. \textbf{11} (2001), no.~3, 694--734.

\bibitem{vereten80}
A.~Yu. Veretennikov, \emph{On strong solutions and explicit formulas for
  solutions of stochastic integral equations}, Mat. Sb. (N.S.)
  \textbf{111(153)} (1980), no.~3, 434--452, 480.

\bibitem{vereten82}
\bysame, \emph{Parabolic equations and {I}t\^o's stochastic equations with
  coefficients discontinuous in the time variable}, Math. Notes \textbf{31}
  (1982), 278--283 (English).

\end{thebibliography}
\bibliographystyle{amsplain}

\end{document}